# Antifouling membranes for oily wastewater treatment: interplay between wetting and membrane fouling

Shilin Huang[1], Robin H. A. Ras[2,3]*, Xuelin Tian[1]*

[1]School of Materials Science and Engineering, Key Laboratory for Polymeric Composite & Functional Materials of Ministry of Education, Sun Yat-sen University, Guangzhou 510006, China

[2]Aalto University, School of Science, Department of Applied Physics, Puumiehenkuja 2, 02150 Espoo, Finland

[3]Aalto University, School of Chemical Engineering, Department of Bioproducts and Biosystems, Kemistintie 1, 02150 Espoo, Finland

**Email:** tianxuelin@mail.sysu.edu.cn, robin.ras@aalto.fi

**Abstract**

   Oily wastewater is an extensive source of pollution to soil and water, and its harmless treatment is of great importance for the protection of our aquatic ecosystems. Membrane filtration is highly desirable for removing oil from oily water because it has the advantages of energy efficiency, easy processing and low maintenance cost. However, membrane fouling during filtration leads to severe flux decline and impedes long-term operation of membranes in practical wastewater treatment. Membrane fouling includes reversible fouling and irreversible fouling. The fouling mechanisms have been explored based on classical fouling models, and on oil droplet behaviors (such as droplet deposition, accumulation, coalescence and wetting) on the membranes. Membrane fouling is dominated by droplet-membrane interaction, which is influenced by the properties of the membrane (e.g., surface chemistry, structure and charge) and the wastewater (e.g., compositions and concentrations) as well as the operation conditions. Typical membrane antifouling strategies, such as surface hydrophilization, zwitterionic polymer coating, photocatalytic decomposition and electrically enhanced antifouling are reviewed, and their cons and pros for practical applications are discussed.

## 1. Introduction

   Oily wastewater, which can cause extensive pollution to water and soil, is produced from various sources, e.g., from oil/gas recovery, metal finishing, mining, transportation and oil refining, et al.[1-15] Some oily substances (e.g., phenols, petroleum hydrocarbons and polyaromatic hydrocarbons) are toxic and can inhibit the growth of plants and animals. To human beings they also bring mutagenic and carcinogenic risks. Direct disposal of the oily wastewater is forbidden by government regulations,[16, 17] and oils in the oily wastewater should be removed to meet the discharge standard (i.e., the highest acceptable concentration of oil/grease in the wastewater is typically in the range of 5-42 ppm depending on the country and location of the platform).[3, 18-

22]

Generally, oily wastewater contains oils in different forms, including free-floating oils, unstable dispersed oils and stable emulsified oils.[11, 13, 17, 23] Unlike free-floating oils (e.g., spilled oils on the ocean), dispersed oils are randomly distributed in water. The dispersed oils have a strong tendency to coalesce and spontaneously evolve into free-floating oils. In contrast, emulsified oils are rather stable due to the presence of surfactants (or surface-active molecules acting similar to surfactants, e.g., asphaltenes in crude oil). Emulsified oils have small droplet sizes, typically smaller than 10 μm.[4, 8, 13, 17, 23, 24] They are commonly found in produced water from oil/gas recovery and metal finishing industries.[11]

Conventional methods to separate oily wastewater include skimming, sedimentation, centrifugation, dissolved gas flotation and biological methods.[16, 17, 21, 23] Though these methods can be used for treating free-floating oils and dispersed oils, most of them are not suitable for treating emulsified oils because the emulsified oils have small droplet sizes, low density difference compared to water (< 150 kg/m$^3$) and high stability.[11, 17, 23, 25, 26] Membrane filtration provides a highly desirable method for treating oily wastewater containing emulsified oils due to its energy efficiency, ease of processing, and low maintenance cost.[14, 18, 23, 27, 28]

Membrane separation of oily wastewater is basically based on two effects, size exclusion (i.e., sieving) and selective wettability.[5-7, 9, 10] The first effect means that the membrane allows water to pass through under an applied pressure while blocks the oil droplets which are larger than the membrane pores. [29] The second effect guarantees that the oil droplets do not wet and permeate the membrane through its selective wetting properties towards water and oil (e.g., hydrophilicity and underwater oleophobicity).[5, 6, 30-33]

Depending on the pore size and separation mechanism, membrane filtration can be divided into microfiltration (MF), ultrafiltration (UF), nanofiltration (NF) and reverse osmosis (RO). [4, 29] Polymers and ceramics are generally used to fabricate filtration membranes. Polymer membranes are relatively cheap, while ceramic membranes have high mechanical strength, high resistance to harsh environments and long lifetime.[34, 35] Other porous materials, such as metal meshes, textile, nanofiber mats and foams can also be used for pretreatment of oily wastewater. [36-45]

Although membranes with different pore sizes and materials are commercially available, they are susceptible to fouling when used for oil/water separation.[46] Fouling leads to continuous decline of flux over time and severely decreases the efficiency of filtration. When the membranes are badly fouled, physical cleaning (e.g., water flush and backflush) or/and chemical cleaning methods have to be performed.[1, 4, 18, 21, 41, 47-50] Fouling leads to higher operation cost and shortens the lifespan of the membrane, impeding the wide applications of membrane technology in oily wastewater treatment.[51]

## 2. Fundamentals of membrane fouling
## 2.1 Classification of fouling

Generally speaking, membrane fouling is caused by complicated interactions between components in the feed solution and the membrane (see Fig. 1a, b for the typical configurations of membrane filtration setup), which are related to their physicochemical properties.[52] Fig. 1c illustrates the change of permeate flux (i.e., volumetric flow rate of permeate per unit of membrane area, $L/m^2 h$) during treatment of oily wastewater using membrane.[53-56] In region I (t$_0$ → t$_1$), when pure water is used as the feed, the flux of pure water $J_0$ depends on the size of the membrane pores, the porosity and the applied pressure (as described by Darcy's law). In region

II ($t_1 \rightarrow t_2$), the oily wastewater is filtrated. The flux $J(t)$ declines overtime due to the fouling of the membrane (at $t_2$ the flux decreases to $J_1$). The membrane fouling can be either reversible or irreversible. The reversible fouling refers to the fouling that can be cleaned up by simple physical methods, such as water flush or backflush.[22, 30, 57, 58] In region III ($t_2 \rightarrow t_3$) of Fig. 1c, when water is reused as the feed to wash the membrane, the flux can be recovered to $J_2$. The fouling that can be recovered by physical cleaning is reversible fouling ($J_{\text{rev}} = J_2 - J_1$). In contrast, the fouling that cannot be recovered by physical cleaning is called irreversible fouling ($J_{\text{irrev}} = J_0 - J_2$).[30] The cleaning up of irreversible fouling requires more intense methods, e.g., using chemicals or applying thermal treatments.[21, 53] Once the irreversible fouling becomes serious, the membrane modules have to be replaced.[14]

Compared to reversible fouling which can be mitigated by optimizing the operation conditions, irreversible fouling is more relevant to the surface chemistry and structure of the membranes. In the following section, the fouling models in wastewater treatment will be discussed.

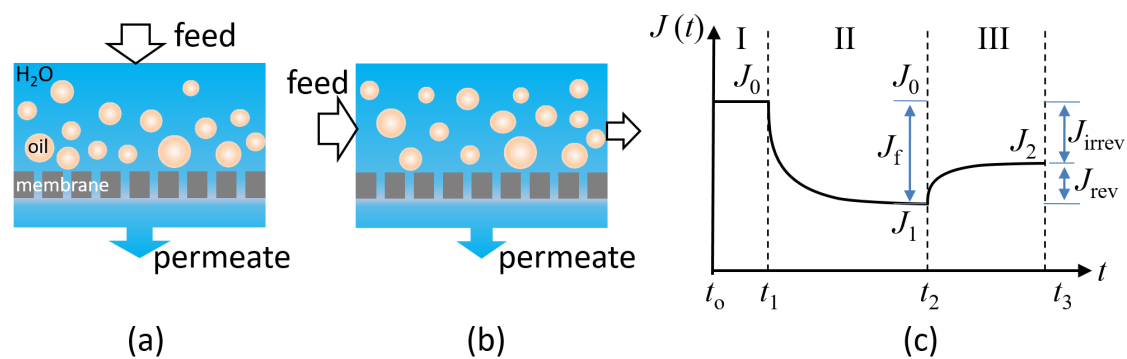

Fig. 1 Schematic illustration of typical membrane filtration modes, including (a) dead-end and (b) crossflow filtration. (c) Permeate flux during filtration of pure water (region I), oily wastewater (region II) and again pure water (region III).

## 2.2 General fouling models
### 2.2.1 Resistance-in-series model

The general form of the resistance-in-series model, given in Eq. (1), can be used to quantify the contribution of each fouling mechanism to the flux decline during filtration:[52, 53, 59-62]

$$J = \frac{\Delta P}{\mu(R_{\text{m}} + R_{\text{rev}} + R_{\text{irrev}})}$$

Eq. 1

where $J$ is the permeate flux, $\Delta P$ is the cross-membrane pressure, and $\mu$ is the viscosity. The resistances $R_{\text{m}}$, $R_{\text{rev}}$ and $R_{\text{irrev}}$ are the hydraulic resistance of the fresh membrane, the hydraulic resistances due to reversible and irreversible fouling, respectively.[62, 63] The reversible fouling resistance $R_{\text{rev}}$ is removable by physical means, e.g., by backflush or switching the feed to pure water.[58] The irreversible fouling resistance $R_{\text{irrev}}$ reflects the deposition of material on the membrane that cannot be removed by physical cleaning.

The resistance-in-series model provides a method to quantify the reversible and irreversible fouling during filtration. $R_{\text{rev}}$ and $R_{irrev}$ normally increase quickly at the beginning of filtration, but slow down during long-time operation. For operation at constant pressure, a steady state (constant $R_{\text{rev}}$ and $R_{\text{irrev}}$) may be reached if there is a balance between the accumulation of foulants and their removal away.[3, 23, 63] Note that $R_\text{rev}$ and $R_\text{irrev}$ are also dependent on the operation conditions (e.g., applied pressure, flow velocity, and physical cleaning methods). Eq. (1) also implies that a higher applied pressure can lead to the increase of permeate flux. However, membrane fouling may become more serious at a higher pressure. Thus, during oily wastewater treatment, it is necessary to optimize the operation conditions in order to obtain a high permeate flux and meanwhile prevent serious membrane fouling.

### 2.2.2 Hermia's fouling models

Membrane fouling depends on the size of foulant (either solid or liquid), foulant-membrane and foulant-foulant interactions. Hermia's fouling models are widely used to describe the flux decline (i.e., fouling) during membrane filtration.[62, 64] These models include complete blocking model, standard blocking model, intermediate blocking model and cake filtration model. In complete blocking model, each foulant particle blocks a pore of the membrane without superimposition upon each other, thus the blocked surface area is proportional to the permeate volume. In standard blocking model, the size of the particle is smaller than the pore diameter. As a result, the pollutant particles can enter the pores and deposit on the pore walls, leading to the reduction of the pore's volume which is proportional to the permeate volume.[62] In intermediate blocking model, it is assumed that not all foulant particles are in direct contact with the pores, but some of them sit on top of others. In the cake filtration model, large amounts of foulant particles accumulate on the membrane and form a cake layer, which creates an additional resistance to the permeate flow. These models predict different decline trends of permeate flux during filtration. They have been employed in analyzing the experimental results in oily wastewater treatment using membranes.[13, 28, 35, 59, 65]

In these fouling models, the foulant particles which enter the pores of the membrane or strongly adsorb on the membrane surface can contribute to irreversible fouling. Otherwise, they can be easily washed away and contribute to the reversible fouling.

For oily wastewater treatment, membrane fouling is expected to be more complicated than the classical Hermia's fouling models. First, oil droplets may wet the membrane surface and the pores. Second, oil droplets accumulating on the membrane can deform and coalescence during filtration. These specific behaviors which significantly influence membrane fouling during oily wastewater treatment will be discussed in details below.

## 3. Fouling mechanisms in oily wastewater treatment
## 3.1 Fouling of membrane by oil

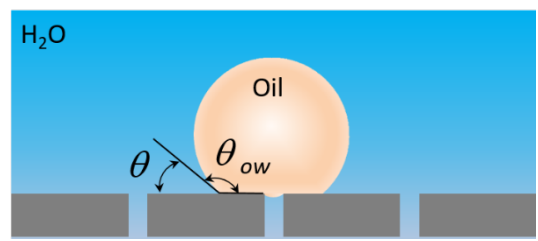

Fig. 2 Illustration of an oil droplet under water sitting on top of a porous membrane.

**Wetting behavior of oil droplet on membrane.** Fig. 2 shows an oil droplet in direct contact with the membrane under water. The membrane has an intrinsic underwater contact angle $\theta_{ow}$ larger than 90º. In this sketch the oil droplet is sitting on top of an idealized pore (i.e., cylindrical pore with vertical sidewalls).[3] The oil droplet can cross the membrane when the applied pressure is larger than the critical pressure ($P_{crit}$). Following Nazzal and Wiesner,[66] $P_{crit}$ can be calculated using Eq. 2 (Eq. 2 is slightly different from that given in the original paper of Nazzal and Wiesner because of a typographical error, as noted by Cumming et al.[67]):

$$P_{crit} = 2\gamma_{ow}\frac{\cos\theta}{r_{pore}}\left[1-\left\{\frac{2+3\cos\theta-\cos^3\theta}{4(\frac{r_{drop}}{r_{pore}})^3\cos^3\theta-(2-3\sin\theta+\sin^3\theta)}\right\}^{1/3}\right]$$

Eq. 2

where $\gamma_{ow}$ is the interfacial tension between oil and water, and $\theta$ is the contact angle measured from the water side, i.e., $\theta = 180° - \theta_{ow}$ (see Fig. 2). $r_{pore}$ and $r_{drop}$ are radii of the pores and oil droplets, respectively.

Several conclusions can be drawn from Eq. 2. First, the underwater oil contact angle $\theta_{ow}$ determines the sign of $P_{crit}$. For $\theta_{ow}$ < 90°, $P_{crit}$ is negative, implying that the oil can wet and fill the pores of the membrane spontaneously even under zero pressure. The filtration may fail because the oil can easily pass through the membrane. Thus, $\theta_{ow}$ should be larger than 90° to obtain successful filtration, and is preferred to be considerable higher since it allows a high transmembrane operation pressure, which is important to increase the permeate flux.

For a given membrane, $\theta_{ow}$ can be predicted using the experimentally verified correlation:[68] $\cos(180° - \theta_{ow}) = \frac{2\gamma_w\cos\theta_w - \gamma_w - \gamma_o + \gamma_{ow}}{\gamma_w - \gamma_o + \gamma_{ow}}$, where $\theta_w$ is the intrinsic water contact angle of the membrane in air, denoting $\gamma_w$, $\gamma_o$ and $\gamma_{ow}$ as water surface tension, oil surface tension and water-oil interfacial tension, respectively. It is evident that $\theta_{ow} > 90°$ can only be fulfilled when $\theta_w < 90°$ ($\theta_{ow} > 90°$ requires $\cos\theta_w$ to be larger than $\frac{\gamma_w+\gamma_o-\gamma_{ow}}{2\gamma_w}$, which is a positive value. Thus $\theta_w$ has to be smaller than $90°$). In another word, a membrane has to be hydrophilic ($\theta_w < 90°$) in order to obtain $\theta_{ow} > 90°$, which is requisite for successful filtration. However, it should be noted that a hydrophilic surface with $\theta_w$ < 90° is not certainly oleophobic underwater. In fact, the boundary with respect to $\theta_w$ between underwater oleophobicity and underwater oleophilicity is normally much less than $90°$. For example, Tian et al. showed that in a hexadecane-water system, a surface is underwater oleophilic when $\theta_w$ is less than 65° and underwater oleophobic when $\theta_w$ larger than 65°.[69]

Second, when $\theta_{ow}$ is > 90°, the critical pressure increases with decreasing pore radius. This means that membranes with smaller pores have higher rejection efficiency to oil droplets.

Lastly, Eq. 2 also indicates that larger droplets have a higher critical pressure, and thus smaller droplets are easier to pass through the pores under pressure. Obviously, if the droplets are smaller than the pore size, the droplets would freely pass through the membrane pores, leading to failure of filtration. For droplets of infinite large size (e.g., an oil film covering the membrane), the critical pressure becomes $P_{crit} = 2\gamma_{ow}\cos\theta/r_{pore}$. [70]

**Oil fouling models.** During filtration of oily wastewater, emulsified oil droplets are carried towards the membrane following the permeate flow and then deposit on the membrane surface. The deposited droplets would block partially the membrane pores at the early stage of filtration (Fig. 3a). With prolonged filtration time, more and more oil droplets accumulate on the membrane surface, leading to the formation of cake layer (Fig. 3b). In crossflow filtration (in such filtration mode the feed flow travels tangentially across the membrane surface, see Fig. 1b), as the crossflow could also carry oil away from the cake layer, a steady cake layer is expected to form once a balance between oil deposition and oil removal is reached, as noted before. Since oil droplets are deformable, the resultant cake layer can be densely packed and shows high resistance to water permeation.[8] Oil droplets in the cake layer are thermodynamically unstable and tend to coalesce.[71] In some experiments, it was found that coalescence led to formation of larger oil droplets which were easier to remove by crossflow.[3, 8] This can be understood by considering the critical penetration pressure of oil droplets. Larger droplets have higher critical pressure (see Eq. 2), thus they are less likely to permeate through the membrane and more probable to be carried away by the crossflow (the critical droplet size beyond which the oil droplet can be carried away by crossflow has been predicted based on the force balance on the droplet).[3] In this respect, coalescence of oil droplets helps to mitigate membrane fouling.

However, if severe pressure is exerted on the oil droplets, they may experience a wetting transition on the membrane (especially for less oleophobic membranes), accompanying significant decrease of their oil contact angles under water. [11] These collapsed oil droplets might coalesce laterally on/within the membrane, and consequently a contiguous oil film forms (Fig. 3c). It is expected that the contiguous oil film forms more easily on membranes which are underwater oleophilic. [8] Such contiguous oil film obviously brings serious membrane fouling.

For droplets smaller than or comparable to the pore size of membrane, droplets could enter or be forced into the pores by the permeate flux. This leads to internal oil fouling within the pores (Fig. 3d), also a type of membrane fouling that is difficult to clean. [3, 8]

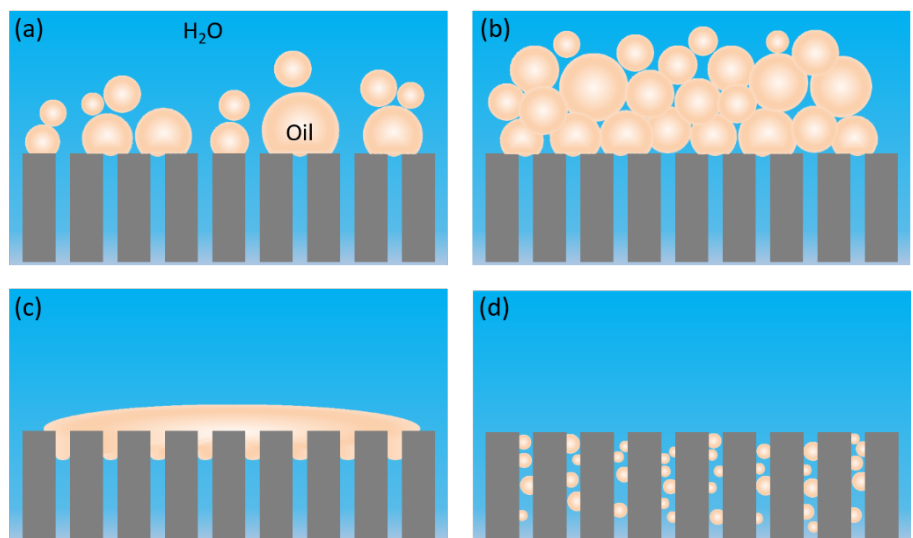

Fig.3 Different oil fouling models in oily wastewater filtration. (a) Oil droplets partially blocking the membrane pores. (b) Cake layer formation on the membrane. (c) Contiguous oil film on the membrane. (d) Oil droplets within the membrane pores.

**3.2 Controlling factors of membrane fouling in oily wastewater treatment**

*Effect of surface chemistry.* As discussed in the previous section, one controlling factor of membrane fouling is the affinity between oil droplets and the membrane under water (i.e., underwater oleophilicity/olephobicity). Oil droplets with high affinity to the membrane can wet the membrane and permeate into the pores, leading to serious fouling. Thus, a poor affinity between oil and membrane under water (i.e., underwater oleophobicity) is a requisite for achieving antifouling membrane. Considering that a membrane with higher hydrophilicity shows higher underwater oleophobicity,[68, 69] hydrophilic membranes can be chosen for antifouling purpose.

***Effect of pore size.*** The pores should be sufficiently small in order to have a good size-sieving effect as well as to prevent "standard pore blocking". From a practical point of view, however, the pores should not be too small, otherwise the membrane resistance to permeate flux would be too high.

***Effect of surface structure.*** The surface roughness influences the membrane fouling on two respects. On the one hand, due to the hydrophilicity of the membrane, water can be trapped in the micro-/nanoscale rough structure. Oil droplets in contact with the membrane are in fact contacting a composite interface with a high portion of water, which could bring extremely low adhesion to oil.[37, 72-76] On the other hand, it is widely reported that increasing the roughness leads to higher fouling tendency due to the accumulation of oil at the valley of the rough surface.[1, 23, 52, 61, 77-80] It seems that the effect of surface roughness on membrane fouling is dependent on the size of oil droplets in respect to the characteristic length of the roughness. If the oil droplets are significantly larger than this characteristic length, an underwater superoleophobic state which decreases fouling tendency can be obtained. On the contrary, if the oil droplets are small compared to the roughness, they could be trapped at the valley of the rough surface, leading to accumulation of oil and membrane fouling.

***Effect of surface charge.*** Oil droplets and membranes can carry surface charges under water. It is generally accepted that if the membrane and the oil droplets have different surface charges, the electrostatic attraction would promote membrane fouling, and vice versa. [58, 79, 81] This electrostatic attraction/repulsion between the oil droplets and the membrane can be estimated based on the classical Derjaguin-Landau-Verwey-Overbeek (DLVO) theory.[82] It is also reported that surface charges can influence membrane fouling by modifying its wettability towards oil droplets.[83]

***Effect of surfactants.*** Surfactants are generally present in oily wastewater.[16, 30, 74, 84, 85] Their influence on membrane fouling during wastewater treatment is multiple.[11, 13, 18, 22, 86, 87] First, surfactants can be adsorbed on/in the membrane and increase its resistance to water permeation (especially for UF and NF membranes since they have small pores), and the surfactant micelles may also block the pores leading to flux decline.[11] In this respect, surfactants act as foulants during membrane filtration.

Second, surfactants can alter the wetting behavior of oil droplets on the membrane.[13] A hydrophilic membrane may become less hydrophilic and more oleophilic upon adsorption of surfactants, and vice visa.[88] This is because the hydrophilic (polar) end groups of surfactants would preferentially adsorb onto the hydrophilic membrane surface, whereas the hydrophobic hydrocarbon chains are likely to be exposed outwards. Consequently, the contaminated hydrophilic membrane becomes prone to be fouled by oils (i.e. showing less antifouling capability). Meanwhile, the membrane would also lose its selective wettability towards oil and water upon surfactant adsorption, which also adversely influence its efficiency in oil-water separation.

Interestingly, a recent study by Schutzius et al.[89] showed that water-soluble surfactants with concentration above the critical micelle concentration (CMC) could impart high underwater oil contact angles (larger than 150º) for a wide range of surfaces, such as glass, aluminum, poly(methylmethacrylate) and poly(vinylidene fluoride), irrespective of their intrinsic wetting properties. They suggested to use such effect (i.e. the detergency effect) for oil/water separation, though the use of high-concentration surfactants may cause environmental concern. [90]

In addition, surfactants decrease the oil/water interfacial tension. This effect facilitates the deformation of oil droplets and their permeation through the pores (since the critical penetration pressure decreases, see Eq. 1), which could adversely influence the separation efficiency.

At last, surfactants (being anionic or cationic) impart charges on the oil droplets. The attractive (or repulsive) electrostatic forces between the membrane and oil droplets would increase (or decrease) membrane fouling tendency, as discussed in the previous section. Lu et al., however, reported an unusual phenomenon: irreversible fouling was alleviated when the charge of surfactant-stabilized oil droplets was opposite to the ceramic membrane during UF.[13, 22] This phenomenon was explained by the synergetic steric effect and demulsification effect.[13] The steric effect meant that at the beginning of filtration, some surfactants were adsorbed on/in the membrane due to electrostatic attractions. The adsorbed surfactants acted as barriers for oil penetration. The demulsification effect meant that, as surfactants became less available to stabilize oil droplets (because some of them were adsorbed on the membrane), oil droplets close to the membrane became unstable and tend to coalesce (demulsify). Because the coalesced droplets were more likely to be rejected by the membrane, the irreversible fouling was alleviated.

*Effect of salts.* Oily wastewater often contains certain amount of salts, which also influences membrane fouling during filtration.[1] First, salts can change the oil-water interfacial tension, influencing droplet deformation and penetration through the membrane.[11] Second, salts may promote droplet coalescence due to electrostatic screening. This also influences membrane fouling. Moreover, antifouling membranes may gradually lose their antifouling property under saline water because of the decomposition and corrosion of hydrophilic components of the membrane.[72, 91] At last, during treatment of saline wastewater the membranes can also be contaminated by salt crystals.[2, 50]

*Effect of operation conditions.* Operation conditions also influence membrane fouling.[1, 17, 34, 63, 90, 92] The filtration module should be designed to have appropriate hydrodynamic conditions to mitigate fouling.[26] For example, the crossflow configuration shows less fouling compared to the dead-end configuration (see Fig. 1a, b for the schematics of these two configurations).[18, 74] Using pulsated feed flows or other perturbations at the membrane surface (e.g., applying continuous or pulsated electric fields) can also efficiently decrease membrane fouling. [34] Since more concentrated oily wastewater is more prone to foul the membrane,[51] it is helpful to perform pretreatments (e.g., using flocculation or pre-filtering to decrease oil concentration) before filtration.[1, 38, 51]

## 4  Methods of testing fouling

Membrane filtration can be performed under a constant pressure,[3, 13, 22, 58] and/or under a constant crossflow rate.[8, 18, 21, 58] During filtration, membrane fouling is reflected by the decline of permeate flow over time. Direct measurement of the permeate flux decline during filtration is a standard method of quantifying fouling. In some filtration processes, the permeate flux is kept constant, and the pressure increases during filtration due to fouling.[8, 11, 63] In this case, the pressure increase can be used to quantify membrane fouling.

To quantify the reversible and irreversible fouling, physical cleaning (e.g., water flush and backflush) is applied to the membrane. The permeate flux (or pressure) which can be recovered by physical cleaning refers to reversible fouling, and the permeate flux (or pressure) that cannot be recovered refers to irreversible fouling. Cyclic filtration processes with interval physical cleaning can be used to test the long-term antifouling performance of the membrane.[22, 58] The membrane showing less flux decline after cyclic filtration is regarded to have a better antifouling performance.

When membranes with sufficient optical transparency in the wet state are used, the fouling dynamics can be directly observed under optical microscope (in-situ method).[3] Ex-situ methods, such as scanning electron microscopy (SEM)[13, 17, 22, 24, 49, 50, 93] and atomic force microscopy (AFM)[2] provide nanoscopic routes to observe the fouled membranes. The macroscopic fouling phenomenon, i.e., oil stain on the membrane, can also be utilized to test the fouling property of the membrane qualitatively.[40, 94]

When an oil droplet approaches and retracts from the membrane surface, a low adhesion force is indicative of low fouling tendency. This adhesion force can be recorded using a force tensiometer,[31, 76, 83, 85, 95, 96] which has a force resolution in the sub-micro-Newton range. For nano-Newton resolution, AFM could be used since it can serve as a powerful method to study the molecular forces between the oil droplet and the solid substrate (e.g., membrane).[82]

## 5  Antifouling strategies
### 5.1 General method: improving surface hydrophilicity

It is normally true that a more hydrophilic substrate is more oleophobic under water.[68, 69] As membranes with higher underwater oleophobicity are more resistive to oil fouling, a general method of preparing antifouling membrane is to improve the hydrophilicity of the membrane.[17, 22, 24, 28, 40, 53, 60, 77, 79, 94, 97-100] In addition, a hydration layer may form on the surface of some hydrophilic materials (such as zwitterionic polymers, polyelectrolytes and polyethylene glycol) under water.[24, 30, 101-105] This hydration layer prevents oil from directly contacting the membrane and thus decreases the fouling tendency. There are various methods to improve the hydrophilicity of the membrane, including surface modification, blending and fabricating nanocomposite membranes. [17, 23, 30, 32, 52, 54, 57, 59, 72, 74, 75, 77, 80, 90, 97, 98, 106-113]

*Surface hydrophilization.* The surface hydrophilicity of a membrane can be increased by surface modification, e.g., introducing hydrophilic polymers or nanoparticles on the membrane surface.[98] Hydrophilic polymers can be introduced to the membrane surface by either surface grafting or coating.[32] Surface grafting requires functional groups on the membrane surface, so that the hydrophilic polymers with reactive groups can be grafted to it via formation of chemical bonds.[30, 54, 74] In comparison, surface coating does not require functional groups on the membrane,[17, 52, 57, 72, 75, 80, 107-110, 114] and the polymers are coated on the membrane by strong physical adsorption. The stability of the coating can be further improved by crosslinking.[40, 115] In addition to hydrophilic polymers, hydrophilic nanoparticles (e.g., metal oxide nanoparticles) are also used to coat the membrane surface to improve the surface hydrophilicity.[97, 111] Surface modification has the advantage of low cost, as it can be easily adopted to modify various commercial membranes.

Generally speaking, surface hydrophilicity is improved by introducing polar groups on the membrane. However, once the polar groups are exposed to air or oil, they tend to orient inward, minimizing the surface energy, see the illustration in Fig. 4 for poly(acrylic acid) (PAA) and poly(sodium, 4-styrenesulfonate) (PAS) decorating surfaces.[116] The surface hydrophilicity may degrade due to such surface reconstruction. Recently, Huang and Wang developed self-cleaning surfaces with stable surface hydrophilicity by coating the surfaces with cellulose nanofibrils (CNFs).[116] The cellulose nanofibrils had a unique isotropic core-corona configuration, which showed a polar corona with uniformly, densely and symmetrically arranged surface carboxyl and hydroxyl groups, and a core with crystalline nanocellulose strands (Fig. 4, left illustration). This configuration enabled large number of polar groups pointing towards the environment, allowing stable surface hydrophilicity.

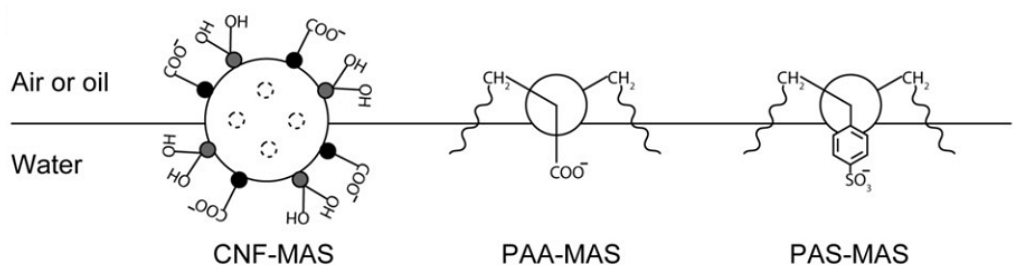

Fig. 4 Illustration of the projections of the molecular structural units of cellulose nanofibril (CNF, left), poly(acrylic acid) (PAA, middle), and poly(sodium, 4-styrenesulfonate) (PSS, right). MAS is the abbreviation of model anionic surface. CNF orients the identical number of carboxyl and hydroxy groups to water and air or oil as a result of its isotropic core-corona configuration, while PAA and PSS orient the ionic groups preferentially to water and the hydrophobic moieties to air or oil owing to the anisotropic configuration. Reproduced with permission from John Wiley and Sons.[116]

Membranes with underwater superoleophobicity (i.e. underwater oil contact angle > 150º) have been widely studied in recent years. Once wetted by water, these membranes can efficiently repel oil and show excellent antifouling property. Underwater superoleophobicity is usually obtained by combining hydrophilic chemical composition and micro/nanoscale roughness on the membrane surface.[19, 37, 74, 76, 101, 115, 117, 118] Due to the intrinsic rough surface structures, membranes are in fact expected to exhibit underwater superoleophobicity once their surfaces are effectively hydrophilized. A number of methods, such as hydrogel coating and salt-induced phase-inversion approach have been employed to fabricate membranes with underwater superoleophobicity.[19, 31, 102, 115, 117, 119, 120] For example, Gao et al. reported a polyionized hydrogel coated copper mesh (underwater oil contact angle ~ 165º), which exhibited ultralow adhesion to viscous crude oils under an aqueous environment and could effectively separate a crude oil/water mixture with high flux and high oil rejection. [100]

***Blending and fabricating nanocomposite membranes.*** Surface modification only imparts a thin hydrophilic layer on the membrane surface (also on the walls of the pores). Therefore, the long-term stability of the resulting hydrophilic surface is relatively poor.[112] This problem can be solved by incorporating hydrophilic materials in the membrane through blending or/and fabricating nanocomposite membranes. For blending, copolymers with hydrophilic components are blended with the membrane matrix.[59, 77, 90] For fabricating nanocomposite membranes, hydrophilic nanoparticles (e.g., metal oxides, graphene oxide, etc.) are dispersed into the membrane matrix for membrane preparation.[23, 47, 52, 56, 94, 98, 111, 113, 121, 122]

**5.2 Zwitterionic coating**

Zwitterions are neutral molecules with equal numbers of positively and negatively charged functional groups. Zwitterionic polymers, which have zwitterionic functional groups in every repeating unit of the polymer, are highly resistant to oil fouling.[32, 54, 75, 81, 105, 109, 110, 123] Their fouling resistance comes from the fact that they superiorly bind water molecules via electrostatically induced hydration.[103] Different from other hydrophilic polymers (e.g., polyethylene glycol, PEG) which can only form a hydration layer via hydrogen bonding, the zwitterionic polymer forms a hydration layer via strong electrostatic interactions due to the strong dipole moments in the zwitterionic units.[104, 105] Molecular dynamics simulations showed that for $-N^+(CH_2)_2SO_3^-$ sulfobetaine there were about 7 water molecules around a sulfonate group and 19 water molecules around a quaternary ammonium group.[124] The tightly immobilized hydration layer at the zwitterionic polymer-water interface has been detected using, for example, sum-frequency-generation vibrational spectroscopy and low-field nuclear magnetic resonance.[103, 104, 125, 126] Wu et al. revealed that there were about 8 water molecules tightly bound with one sulfobetaine zwitterion unit for poly(sulfobetaine methacrylate) modified surface.[104] The tightly bound water layer at the surface of zwitterionic polymer forms a barrier for oil fouling. For more details about the antifouling mechanisms of zwitterionic polymers, readers are referred to Ref. [127].

He et al. grafted zwitterionic poly(2-methacryloyloxyethyl phosphorylcholine) brushes on solid substrates, and the resulting surface was underwater superoleophobic.[103] As the surface was rather flat, the underwater superoleophobicity was not due to surface roughness, but solely due to the intrinsic hydration layer on the zwitterionic polymer surface. The resultant surface exhibited complete oil repellency when it was wetted by water. In the dry state, as expected, the zwitterionic surface could be easily fouled by oil. However, once the fouled surface was immersed in water, the oil spontaneously detached from the surface, see Fig. 5a. Such intrinsic oil repellency of zwitterionic polymer under water is rather unique. In contrast, most underwater superoleophobic surfaces with micro/nanoscale hierarchical structures are difficult to maintain the underwater superoleophobicity once contaminated by oil.[103, 116] Underwater superolophobic surfaces based on polyelectrolyte grafting also cannot be re-wetted by water if they are pre-wetted by oil, because when in contact with oil the polyelectrolyte surfaces reorient their ionic groups inwards to lower the surface free energy, leading to the loss of surface hydrophilicity.[82, 103]

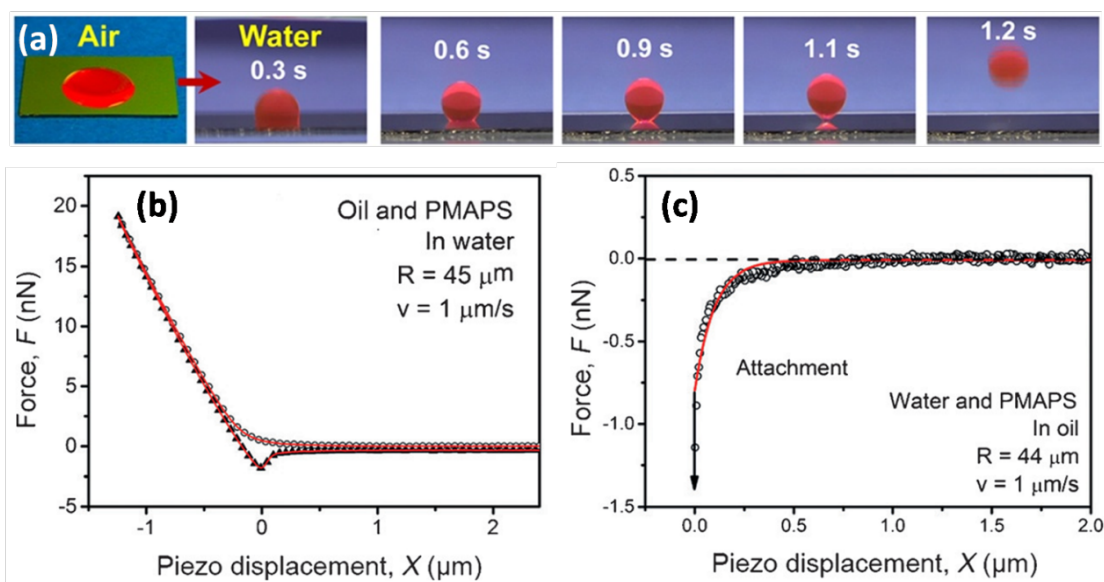

Fig. 5 (a) Time-lapse photos taken after immersion of 60 μL canola oil fouled substrate in water. The substrate is grafted with zwitterionic poly(2-methacryloyloxyethyl phosphorylcholine) brushes. Reprinted with permission from [103]. Copyright 2015 American Chemical Society. (b,c) Measured (open symbols) and calculated (red curves) interaction forces of zwitterionic poly(3-

[dimethyl(2-methacryloyloxyethyl) ammonium] propanesulfonate) (PMAPS) with an oil droplet in water (b) and with a water droplet in oil (c). Positive and negative interaction forces represent repulsive and attractive forces between the droplet and substrate, respectively. The arrow in (c) indicates attachment of the water droplet on PMAPS surface. Open circles are force data measured during approach, and solid triangles are force data measured during retraction. Adapted with permission from John Wiley and Sons.[82]

Shi et al. measured the force between an oil droplet and a zwitterionic polymer surface under water using AFM (Fig. 5b),[82] and no obvious attraction and adhesion forces were detected when the oil droplet approached and retracted from the surface, respectively. Surprisingly, when a water drop approached the zwitterionic polymer surface under oil, as shown in Fig. 5c, a long-range "hydrophilic" attraction was observed. It was attributed to a strong dipolar interaction between the water droplet and the zwitterionic polymer surface.

As a promising strategy, zwitterionic polymers have been used to fabricate membranes with complete resistance to irreversible fouling (either by blending or grafting).[14, 103] Kaner et al. [14] showed that increasing the zwitterionic content in the additive copolymer (containing zwitterionic components) did not always result in improved membrane performance. During membrane formation via non-solvent induced phase separation, the additive copolymer with high zwitterion content (51-52 wt%) led to macrophase separation from the membrane matrix, leading to a poor membrane performance. On the contrary, with appropriate copolymers that contained 18-19 wt% zwitterionic monomer, membranes with high permeate flux and remarkable fouling resistance were obtained even with very small amounts of additive copolymer.[14]

### 5.3 Combining fouling-resistant and fouling-release mechanisms

Several groups have reported enhanced antifouling property of membranes by combining fouling-resistant and fouling-release mechanisms.[18, 55, 95, 128-132] In the fouling-resistant mechanism, water molecules are tightly bound to the hydrophilic surface and form a hydration layer, preventing oil from contacting the surface. In the fouling-release mechanism, the surface is covered by low-surface-energy fluorine atoms, which reduces the adhesion energy between oil and the surface and facilitates the release of adsorbed oil. [131, 132]

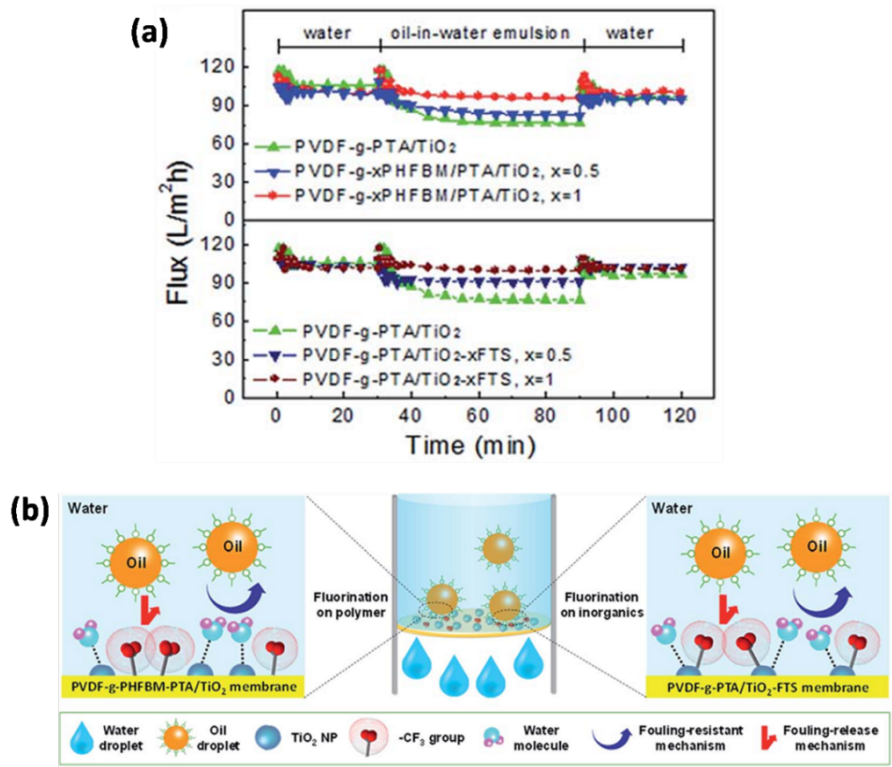

Fig. 6 (a) Time-dependent permeate fluxes for hybrid membranes in three-stage filtration: 0.5 h water filtration, 1 h oil-in-water emulsion filtration and 0.5 h water filtration after rinsing. Top: the -$CF_3$ groups are grafted in the PVDF matrix. Bottom: -$CF_3$ groups are introduced on the $TiO_2$ particles. The results show that the antifouling property of the membrane is improved by increasing -$CF_3$ content. (b) Schematic of collaborative defense mechanisms for heterogeneously constructed hybrid membranes. Hydrophilic $TiO_2$ components contribute to the fouling-resistant mechanism, -$CF_3$ groups contribute to fouling-release mechanism. Adapted with permission from Royal Society of Chemistry. [132]

Zhao et al. fabricated polyvinylidene fluoride (PVDF) based hybrid membranes using in-situ biomimetic mineralization and non-solvent induced phase separation.[132] The resulting surfaces, as reported by the authors, had both inorganic hydrophilic components $TiO_2$ (facilitating formation of hydration layer) and organic low-surface-energy components (-$CF_3$). The low-surface-energy components (-$CF_3$) were either anchored on the $TiO_2$ surface or grafted to the PVDF matrix. The surface energy decreased with -$CF_3$ content, while the oil fouling resistance was significantly enhanced, see Fig. 6a. The proposed mechanism is shown in Fig. 6b. The hydrophilic $TiO_2$ contributed to the fouling resistance mechanism, while the -$CF_3$ groups on the membrane surface contributed to the fouling release mechanism.

Wang and Lin integrated low-surface-energy perfluoroalkyl functional groups into membranes with chitosan based hydrogel surface.[95] The resulting membranes showed excellent anti-fouling property when treating crude-oil-in-water emulsions, as long as the perfluoroalkyl functional groups were not excessive on the surface.[95] It was postulated that the low-surface-energy moieties in the hydrogel surface promoted the release of foulants by local hydrodynamic perturbation. However, as noted by the authors, there is still no direct proof regarding such mechanism either by experiments or simulations.[95] Since overabundance of low-surface-energy functional groups has a negative effect on the antifouling property, there might be an intermediate concentration of low-surface-energy functional groups at which the best antifouling property can be obtained. However, such optimized condition has not been systematically studied yet.

While the combination of fouling-resistant and fouling-release mechanisms has been qualified as a potential antifouling strategy, further understanding and verification of such antifouling mechanism is still needed. It is also important to develop a criterion for designing such antifouling surfaces if possible.

**5.4 Photocatalytic cleaning**

Generally speaking, if the membrane is irreversibly polluted by oil or other organic compounds in the feed solution, chemical cleaning or high-temperature decomposition should be applied to clean the membrane.[48-50] This cleaning process takes extra time and operation costs. Using photocatalytic nanoparticles such as $TiO_2$ and ZnO, it is possible to prepare membranes with self-cleaning properties.[84] Under UV light or sunlight, photocatalytic nanoparticles are able to generate highly reactive species like superoxide anions and hydroxyl radicals to decompose the organic contaminants.[43, 133] This provides a remote-controlled and non-stop antifouling strategy.[43, 48, 133-136]

For example, Li et al. fabricated a multifunctional underwater superoleophobic porous membrane by growing anodized hierarchical $TiO_2$ nanotubes on the surface of porous titanium.[84] They demonstrated that once the membrane was contaminated by octadecyltrimethoxysilane, it lost its superhydrophilicity. However, after the illumination of UV light the superhydrophilicity was recovered. In addition to the self-cleaning property, the membranes with photocatalytic functionality had the ability to decompose toxic water-miscible organic molecules when water flowed through the membrane.[84]

As this strategy requires illumination of UV light or sun light,[133] the filtration module needs some special designs, e.g., using transparent windows for light transmission. It is also important to ensure that the polymer membranes themselves are not decomposed by the photocatalytic activity.

**5.5 Electrically enhanced antifouling**

During filtration, the foulant particles (e.g., oil droplets and other foulants) flow to the membrane and form the cake layer. For charged particles, this convective flow can be compensated by applying an electric field, which drives the charged particles away from the membrane and prevents the formation of cake layer. The effect of using an electric field to change the trajectories of charged particles is called electrophoresis, which has been used to mitigate membrane fouling during filtration of wastewater. The electric field can be applied either across the membrane, or using the membrane as an electrode. [137]

For example, Geng and Chen developed antifouling tubular $Al_2O_3$ microfiltration membranes, with the inner layer modified by Magnéli $Ti_4O_7$ which was conductive.[138] The resulting conductive inner layer of the membrane was connected to a direct current (DC) electric field and acted as anode. Meanwhile, a stainless steel wire located at the center of the tubular membrane acted as cathode (see Fig. 7 for the schematic of the electrically-assisted membrane filtration module). The model oily wastewater was a peanut-oil-in-water emulsion stabilized by hexadecyltrimethylammonium bromide (CTAB), thus the oil droplets were positively charged. Thanks to the electrophoresis, the antifouling performance as well as the permeate quality were significantly improved when the electric field was applied.

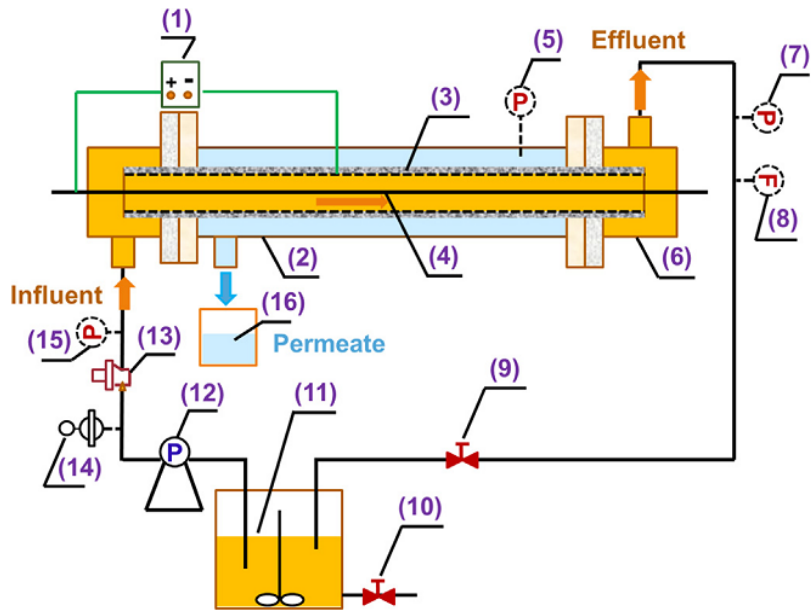

Fig. 7 Schematic diagram of the electrically-assisted antifouling filtration process. (1) DC power supply, (2) annular permeate compartment, (3) inner layer modified $Ti_4O_7/Al_2O_3$ composite membrane, (4) wire electrode, (5) permeate side pressure gauge, (6) retentate compartment, (7) retentate side pressure gauge, (8) flowmeter, (9) pressure control valve, (10) discharge valve, (11) feed solution reservoir, (12) metering pump, (13) back pressure valve, (14) pulsation damper, (15) inlet pressure gauge, and (16) permeate reservoir. Reproduced with permission from Elsevier.[138]

Apart from electrophoresis, the electrochemical reactions can also be used to improve antifouling performance during filtration. Li et al. adopted a coal-based carbon membrane as the anode for treating sodium dodecyl sulfate (SDS) stabilized fuel-oil-in-water emulsion. As the oil droplets were negatively charged, it was expected that the oil would easily foul the membrane which acted as anode. However, the antifouling performance of the filtration system was improved under an electric field. This unexpected phenomenon was believed to be related to the electrochemical reactions taking place at the anode (membrane). The reactive intermediates (e.g., ·OH, $HO_2$·, and $H_2O_2$) on the membrane surface efficiently decomposed and removed oil droplets from the surface.[93]

The electrically enhanced antifouling strategy avoids use of chemicals, consumes low energy and is straightforward for handling. Yet, it has the following issues. First, the electrophoretic mobility of charged particles would be significantly reduced when the feed contains salts (due to electrostatic shielding).[138] This restricts the wide application of electrically enhanced antifouling strategy in oily wastewater treatment since many oily wastewater streams contain salts.[137] Second, it is necessary to finely control the applied voltage for the antifouling mechanism based on electrochemical reactions. Otherwise if the voltage is too high, bubbles can form on the surface of membrane and block the pores, leading to decrease of the permeate flux.[93, 137] At last, if oppositely charged particles are present in the oily wastewater, the charged membrane might attract the particles resulting in unwanted blocking of membrane pores.

**5.6 Hydrophilic dynamic membranes**

Typically, improving the hydrophilicity of membranes requires complicated physical or chemical processes, e.g., surface modification, blending and fabricating nanocomposite membranes.

Alternatively, it can be easily realized by using hydrophilic dynamic membranes. A hydrophilic dynamic membrane is formed by in-situ filtering a coating solution containing either inorganic or organic hydrophilic particles through a supporting membrane.[139] The resulting deposited layer of hydrophilic particles (Fig. 8a) acts as a hydrophilic filtration membrane, which isolates pollutants and protects the supporting membrane from fouling (Fig. 8b). The dynamic membrane shows additional resistance to permeate flux depending on the size of particles (larger particles form dynamic membranes with smaller resistances).[58] As the particles in the dynamic membrane are not chemically connected to each other or to the supporting membrane, they can be easily removed by backwash (Fig. 8c). Therefore, the fouling on the dynamic membrane is reversible. Moreover, after backwash a fresh dynamic membrane can be regenerated by depositing another hydrophilic particle layer. In a word, dynamic membranes have the benefits of simple preparation, easy removal and regeneration. [58, 139]

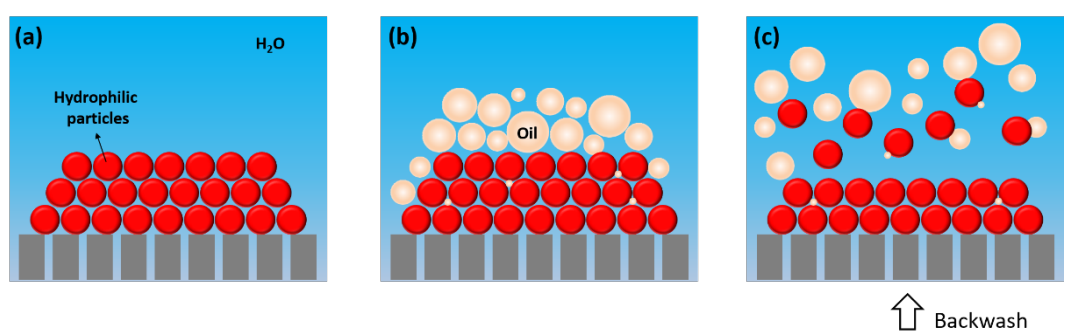

Fig. 8 Hydrophilic dynamic membranes for antifouling. (a) The dynamic membrane, i.e., the deposited layer of hydrophilic particles (red spheres), can be fabricated by filtering a coating solution containing hydrophilic particles through a supporting membrane. (b) The hydrophilic dynamic membrane traps foulants (e.g., the oils, indicated by yellow spheres) and protects the supporting membrane from fouling. (c) The particles and foulants in the dynamic membrane can be easily removed by backwash (indicated by arrow). Re-drawn from Ref. [58].

Under a dead-end filtration condition, Lu et al. used hydrophilic $Fe_2O_3$ particles with an average particle size of 780 nm to fabricate dynamic membrane on an ultrafiltration ceramic membrane.[58] The use of relatively large particles for the dynamic membrane avoided pore blocking on the supporting membrane and guaranteed a small resistance of the dynamic membrane. If $Fe_2O_3$ particles were pre-coated on the supporting ceramic membrane before treating the oil-in-water emulsion, the fouling of the ceramic membrane was significantly reduced. The authors also pointed out that at a neutral pH condition, the electrostatic attractions between the membrane and the $Fe_2O_3$ particles helped to stabilize the dynamic membrane. However, at an alkaline condition (e.g., pH=8), the dynamic membrane and the ceramic membrane showed repulsive forces, facilitating the removal of fouled $Fe_2O_3$ particles by alkalescent water. This mild cleaning condition avoided the use of strong alkaline (pH > 10) and high temperature (~ 80 °C) backwash which could cause severe corrosion to the filtration system.

The strategy using dynamic membranes for antifouling also has some drawbacks. It increases the resistance to permeate flux and requires more investments on the equipment (e.g., the reservoir containing particle solutions should be installed). Moreover, extra efforts should be made to collect, clean and recycle the polluted particles which are washed away by backwash. It is also necessary to optimize the operation conditions for deposition and filtration, in order to have a stable dynamic membrane which prevents oil from contacting the supporting membrane.

**5.7 Magnetic Pickering emulsions for fouling-free separation**

Dudchenko et al. pointed out that underwater superoleophobic membranes did not completely resist fouling under realistic conditions, especially when the oil concentration in the oily wastewater was high.[140] They suggested using Pickering emulsions to decrease fouling during UF. In this strategy, magnetic nanoparticles (diameter ~ 600 nm) were mixed with the oily water to form a Pickering emulsion (oil droplets were stabilized by nanoparticles). The nanoparticles located at the droplet surface efficiently prevented oil droplets from contacting the membrane and thus mitigated membrane fouling by oil. Because the UF membrane had a small pore size, only water passed the membrane while the emulsified oils and nanoparticles were blocked. When this strategy was used to treat oily water with large quantity of crude oil (10%), a minimal fouling tendency was observed. Thanks to the magnetic property of the magnetic nanoparticles, continuous oil-water separation may be possible following the proposed procedures in Fig. 9. [140]

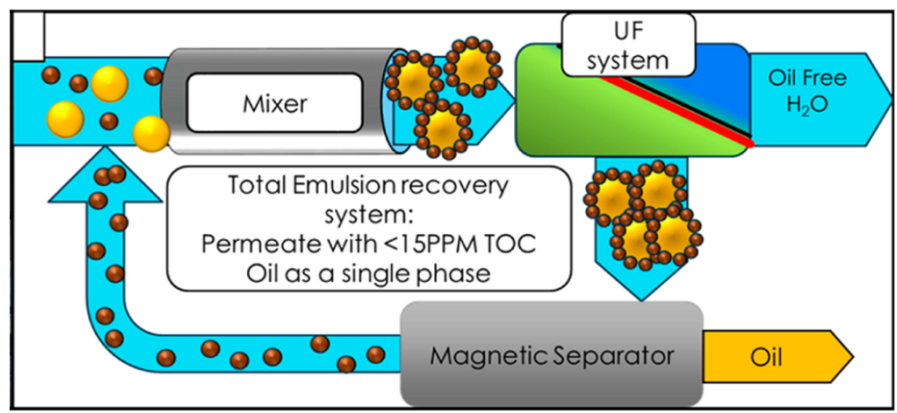

Fig. 9 Complete oil emulsion treatment system: starting from the top left, oily water enters a mixer with $Fe_3O_4$ particles to form a Pickering emulsion; the Pickering emulsion enters the UF system, where an oil-free permeate stream and a concentrated Pickering emulsion stream are produced; the concentrated Pickering emulsion is passed through a magnetic separator, which separates water and $Fe_3O_4$ particles from oil, producing an oil stream and $Fe_3O_4$ particle slurry that is reused for the formation of a new Pickering emulsion (brown dots are $Fe_3O_4$ particles and large yellow dots are oil droplets). Adapted with permission from [140]. Copyright 2015 American Chemical Society.

This antifouling strategy requires that the nanoparticles have a high affinity to the oil-water interface which can be quantified by the detachment energy $U_d$ (the energy required to remove the particle from the interface):[141]

$$U_d = \pi R^2 \gamma (1 \pm \cos\theta_w)^2,$$

(2)

where $R$ is the radius of the particle, $\gamma$ is the interfacial tension between oil and water, and $\theta_w$ is the contact angle of the nanoparticles at the oil-water interface measured from the water side. The sign in the bracket is negative (or positive) if the particle is removed from the interface to the water (or oil) phase. Eq. 2 implies that a high detachment energy demands a high oil-water interfacial tension and a contact angle near 90° (if $\theta_w > 90°$, the particles may detach from the oil side).

If surfactants are present in the system, the interfacial tension $\gamma$ decreases leading to a low detachment energy. In this case, it is difficult to fabricate stable Pickering emulsions. Therefore, the antifouling strategy based on magnetic Pickering emulsions only works effectively in the absence of surfactants. In addition, in order to recover and reuse the magnetic particles, complex post-processing (e.g., magnetic separation and wicking steps) is required,[140] which would increase the operation costs.

**5.8 Other antifouling strategies**

*Liquid-based gating mechanism for antifouling membranes.* Hou et al. proposed an antifouling strategy based on liquid-based gating mechanism (Fig. 10).[142] The pores of the membrane were infused by a low-free-energy liquid, which completely sealed the pores and formed a coating layer. A transport liquid (or gas) which had a lower affinity to the membrane must deform the pore-filling liquid interface in order to enter and penetrate the pores. The critical pressure required for the penetration depended on the interfacial tension between the infusion liquid and the transport liquid. As different transport liquids had different critical pressures, it was possible to separate the liquids by adjusting the operation pressures. In addition, because the transport liquids were only in contact with the infusion liquid, the solid membrane was not fouled by the transport liquids. This anti-fouling strategy was successfully applied to separate air/water/oil mixtures.[142] If such liquid-infused membrane is used to treat oily wastewater, it is not clear whether the infusion liquid would bring undesired effect such as contaminating the filtrate (i.e., water). Moreover, it might be difficult to treat oily wastewater containing surfactants or other organic compounds, because they could alter the surface wettability of the membrane by adsorption and influence the infusion of the gate-forming liquid.

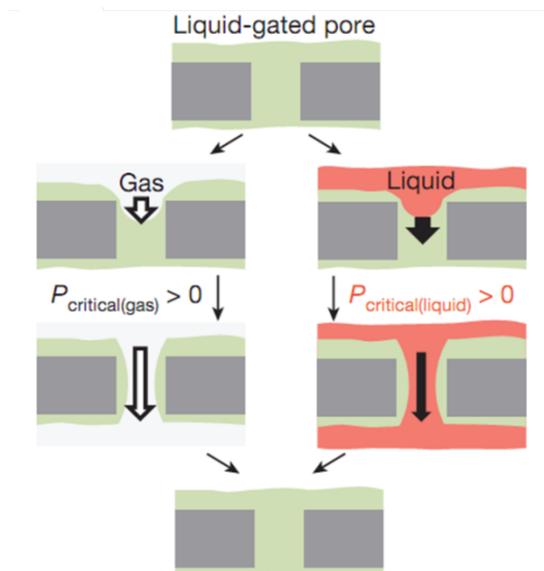

Fig. 10 Liquid-based gating mechanism. If the pore is filled with a stably held liquid (green), flow of gas and liquid will be gated by pressure-induced deformation of the gating liquid surface. When the pressure is higher than the critical pressure ($P_{critical}$), the pores are in the open state. When the pressure is released, the non-fouled pores return to their original liquid-filled state. The liquid-based gating mechanism allows selective, responsive, tunable and antifouling multiphase transport. Reprinted by permission from Macmillan Publishers Ltd: Nature, [142] copyright 2015.

*Detergency effect.* Schutzius et al. found that it was possible to obtain an underwater superoleophobic state regardless of the substrate material by introducing high-concentration surfactants (beyond CMC) into water (i.e. the detergency effect).[89] This effect might be used to decrease oil fouling on membranes during oil-water separation, with no requirement of surface micro-/nanotexturing or chemical modification of the membrane.[89] However, large amount of surfactants have to be present in the wastewater (for example, the CMCs for three typical surfactants, SDS, CTAB and Triton X-100 are 2566 mg/L, 328 mg/L and 155 mg/L, respectively,[71] which are comparable to the oil content in produced water (100-1000 mg/L)).[23] This limits practical application of this method since surfactants are harmful to the aquatic ecosystem.[90]

The antifouling strategies presented above as well as their working principles, advantages and disadvantages are summarized in Table 1.

Table 1. Summary of the antifouling strategies.

| Strategy | Working principle | Advantage | Disadvantage |
|---|---|---|---|
| **Improving surface hydrophilicity** | More hydrophilic substrates are more oleophobic under water and show more robust antifouling performance. | Most widely explored, relatively simple and of low cost. No special requirement for the filtration setup. | Long-term stability of surface hydrophilicity may be limited. |
| **Zwitterionic coating** | Strong binding of water molecules to zwitterionic coating results in a robust hydration layer and intrinsic oil repellency. | Large resistance to irreversible oil fouling. | Chemistry complexity may limit its large-scale application. |
| **Combining fouling-resistant and fouling-release mechanisms** | It is postulated that low-surface-energy groups in a hydrophilic surface can promote the release of foulants. | Promising antifouling performances observed in experiments. | Mechanism needs to be verified and criterion for designing such antifouling surfaces is unclear. |
| **Photocatalytic cleaning** | Photocatalytic nanoparticles generate highly reactive species under UV light or sunlight to decompose organic contaminants. | Possible to decompose water-miscible organic molecules when water flows through the membrane. | Requiring special design of the filtration module for light transmission. Not applicable in systems containing high contents of oil. |
| **Electrically enhanced antifouling** | Electrophoresis prevents charged oil droplets approaching membrane. Electrochemical reactions generate reactive intermediates to decompose and remove oil droplets from surface. | Avoiding use of chemicals, consuming low energy and straightforward for handling. | Sensitive to salt. Possible bubbling and pore blocking. |
| **Hydrophilic dynamic membranes** | Deposited layer of hydrophilic particles acts as a hydrophilic filtration membrane, which isolates pollutants and protects the supporting membrane from fouling. | Simple preparation, easy removal and regeneration. | The resistance to permeate flux is relatively high. It also requires more investments on the equipment, and extra efforts to collect, clean and recycle the polluted particles. |
| **Magnetic Pickering emulsions** | Nanoparticles absorbed at the oil droplet surface efficiently prevent oil droplets from contacting the membrane and thus mitigated membrane fouling by oil. | Able to treat oily water with large quantity of crude oil (10%). | Only works effectively in the absence of surfactants. Reusing the magnetic particles requires complex post-processing. |
| **Liquid-based gating mechanism** | Pores of membrane infused by a low-free-energy liquid can be open or closed based on the transmembrane pressure. | Efficiently preventing contact between transport liquid and membrane thanks to the infusion liquid. | Filtrate may be contaminated by infusion liquid. Might be problematic when treating oily wastewater containing surfactants and organic compounds. |
| **Detergency effect** | Underwater superoleophobic state is obtained with the aid of surfactant (concentration > CMC). | No requirement of surface micro-/nanotexturing or chemical modification of the | Large amounts of surfactants have to be present in the wastewater. |

# 6  Outlook

In addition to oil foulant, real-world oily wastewater contains various other kinds of foulants, such as organic foulants, inorganic foulants and biofilms.[18, 21, 50, 143] It is challenging to prevent fouling by these foulants at the same time. Property of the membrane (e.g., surface hydrophilicity and structure) and the wastewater (e.g., composition and concentration), configuration of the filtration module, and operation conditions are all relevant to membrane fouling during filtration. Except for the antifouling strategies introduced above, it is also helpful to apply pretreatment (e.g., flocculation, pre-filtering, etc.) to the oily wastewater before filtration, [38, 51, 111] or combine different treatment techniques to mitigate membrane fouling.[138, 139]

In the following we present the outlooks on the treatment of oily wastewater using membranes.

(1) Effective and reliable anti-fouling strategies still need to be explored, and some existing antifouling mechanisms need to be further clarified. For example, the combination of fouling-resistant and fouling-release mechanisms has been reported to be an effective antifouling strategy. However, its working principle and design criterion are far from known.

(2) Surfactant adsorption may alter membrane wettability towards water and oil, and thus degrades its antifouling property as well as efficiency in oil-water separation. As surfactants (or similar organic matters) are omnipresent in oily wastewater, it is highly important and also a big challenge to develop new strategies to eliminate the adverse effect of surfactant adsorption, or to develop novel membranes that can resist/reduce surfactant adsorption.

(3) Economic methods should be developed to prepare durable antifouling membranes. Although surfaces coated by zwitterionic polymers or cellulose nanofibrils have shown excellent antifouling property, the durability of such coatings and their large-scale fabrication method remain unsolved.

(4) Except for the surface chemistry, the surface geometry of the membrane also plays an important role in the antifouling property of the membrane.[144] Existing reports mostly studied the effect of pore size and surface roughness on the antifouling property (e.g. to enhance underwater oleophobicity), but it is not clear how specific geometrical structures (e.g. morphologies of pores) affect the oil-membrane adhesion as well as dynamic detachment of oil from the membrane surface. It is of great importance to identify the effect of specific membrane structure on the antifouling properties of membranes.

(5) In the experimental systems where surfactants and salts are involved, the underwater wetting properties of the membrane (e.g., oil contact angle under water) should be measured in water that contains the same amounts of surfactants and salts, since the surfactants and salts can influence the wetting behaviors significantly. Moreover, as the charges of membranes and oil droplets influence membrane fouling, the Zeta potentials of the membranes and oil droplets need to be characterized in order to specify the contribution of charges on fouling.[17, 24, 55, 112]

(6) Intelligent and responsive materials deserve more attention for developing antifouling membranes. For example, Ngang et al. fabricated PVDF/silica-poly(*N*-isopropylacrylamide) membranes which were thermo-responsive. It was found that by temperature swing the irreversible fouling on the membrane could be reduced.[145] It is interesting to investigate whether such intelligent membranes that respond to different stimuli (e.g. photo-, electro-, and magneto-responsive membranes) could bring novel solutions for antifouling purpose.

(7) It is important to develop antifouling membranes that can survive under harsh conditions, such as wastewater with high salinity. For example, during treatment of highly saline wastewater, the hydrophilic coatings on the membrane surface may easily decompose.[72, 91] Inspired by seaweed, this problem can be solved by using alginate as the membrane or coating material.[72, 91] It is believed that similar bioinspired strategies will play important roles in treating oily wastewater using membranes under harsh conditions.

**Acknowledgments**

This work is supported by the "One Thousand Youth Talents" Program of China, and "One Hundred Talents" Program of Sun Yat-sen University. R.H.A.R. acknowledges the European Research Council for funding the Consolidator Grant SuperRepel (grant agreement no 725513).